# Flexible and biocompatible microelectrode arrays fabricated by supersonic cluster beam deposition on SU-8

**Mattia Marelli**[1,4], **Giorgio Divitini**[1,5], **Cristian Collini**[2], **Luca Ravagnan**[1], **Gabriele Corbelli**[1], **Cristian Ghisleri**[1], **Antonella Gianfelice**[1], **Cristina Lenardi**[3], **Leandro Lorenzelli**[2], **and Paolo Milani**[1,6]

[1] Dipartimento di Fisica and CIMAINA, Università degli Studi di Milano, Via Celoria 16, I-20133 Milano, Italy
[2] FBK-irst, Via Sommarive 18, 38050 Povo, Trento, Italy
[3] Dipartimento di Scienze Molecolari Applicate ai Biosistemi and CIMaINa, Università degli Studi di Milano, Via Trentacoste 2, I-20134 Milano, Italy

E-mail: pmilani@mi.infn.it

**Abstract.** We fabricated highly adherent and electrically conductive micropatterns on SU-8 by supersonic cluster beam deposition (SCBD). This technique is based on the acceleration of neutral metallic nanoparticles produced in the gas phase. The kinetic energy acquired by the nanoparticles allows implantation in a SU-8 layer, thus producing a metal-polymer nanocomposite thin layer. The nanocomposite shows ohmic electrical conduction and it can also be used as an adhesion layer for further metallization with a metallic overlayer. We characterized the electrical conduction, adhesion and biocompatibility of microdevices obtained by SCBD on SU-8 demonstrating the compatibility of our approach with standard lift off technology on 4'' wafer. A self-standing and flexible Micro Electrode Array has been produced. Cytological tests with neuronal cell lines demonstrated an improved cell growth and a spontaneous confinement of cells on the nanocomposite layer.

---

[4] Present address: Microsystems Laboratory, École Polytéchnique Fédérale de Lausanne (EPFL), Lausanne 1015, Switzerland
[5] Present address: Department of Materials Science and Metallurgy, University of Cambridge, Pembroke street, CB2 3QZ, Cambridge, United Kingdom
[6] Author to whom correspondence should be addressed.

Flexible and biocompatible microelectrode arrays fabricated by SCBD on SU-8

**1. Introduction**

The use of SU-8 photoresist as a structural material for the fabrication of micro electromechanical devices for biological applications (bioMEMS) is gaining a rapidly increasing attention because of its excellent lithographic properties [1, 2], allowing the inexpensive mass-production of high-aspect ratio structures for applications in microfluidics, drug delivery, neural recording and stimulation [3-5]. Compared to silicon, SU-8 offers several advantages such as high mechanical strength and flexibility, chemical stability, high compatibility with polymeric packaging and encapsulation [6, 7].

The fabrication of complex and reliable implantable biomedical microdevices for *in vivo* and *in vitro* neural applications requires the integration of electrical conduction capabilities with appropriate mechanical properties and biocompatibility [4, 5]. Silicon microneedles and microelectrode arrays for neural and extracellular recording suffer from the rigidity and brittleness of this material, whereas SU-8 has demonstrated to be a superior alternative in terms of flexibility and integration capability especially for the realization of neural interfaces for peripheral nerve monitoring [1].

SU-8 devices need metal patterns in order to handle electrical signals for recording and actuation tasks. Those patterns are usually produced during multi-step fabrication processes by sputtering and/or electroplating of gold films on Cr layers to improve adhesion [8]. Nevertheless, metallization of SU-8 and, in general, of polymeric materials used in bioMEMS represents a challenge due to the lack of large-scale microfabrication methods capable of providing good adhesion of metallic patterns on polymeric substrates, while avoiding polymer damage [7, 9-12].

Here we show that supersonic cluster beam deposition (SCBD) is an effective technique to fabricate highly adherent metallic microstructures and micro electrode arrays on SU-8: in our approach neutral metallic nanoparticles are accelerated by a supersonic expansion to gain enough kinetic energy to penetrate beneath the polymer surface, thus forming a continuous and spatially defined polymer-nanoparticle nanocomposite layer [13]. The microfabricated nanocomposite can be used as an electrically conducting path or as an adhesion layer for subsequent metallization. SCBD has been also used to fabricate microresistors and a self-standing micro electrodes array (MEA) device prototype on SU-8 to be used for *in vitro* experiments on neural cells.

**2. Supersonic cluster beam deposition (SCBD)**

A seeded supersonic beam of palladium clusters is produced using a deposition apparatus equipped with a pulsed microplasma cluster source (PMCS) [14] (figure 1a). The PMCS operates through the ablation of a metallic target by an argon plasma jet, ignited by a pulsed electric discharge [15]. After ablation, metallic atoms thermalize with the inert gas and condense to form clusters; the inert gas-cluster mixture then expands through the nozzle in an expansion vacuum chamber forming a seeded supersonic beam [14]. A set of aerodynamic lenses [16] mounted after the PMCS nozzle is used to obtain a highly collimated cluster beam (with a divergence below 50 mrad) (figure 1a). The cluster implantation takes place on substrates

Flexible and biocompatible microelectrode arrays fabricated by SCBD on SU-8

intercepting the supersonic beam in a second differentially pumped chamber separated from the expansion chamber by an electroformed skimmer (figure 1a). Neutral Pd clusters are accelerated by the supersonic expansion towards the SU-8 substrate with a kinetic energy ($E_k$) of approximately 0.5 eV/atom (the cluster speed is nearly 1000 m/s$^2$) [16]; it should be noted that this is about four orders of magnitude less than the typical kinetic energies for ion beam implantation in polymers [17]. Considering the atomic deposition rate $N_{atm}$ (number of atom reaching a surface unit area per second) we can calculate the surface power density for the cluster implantation: $P_{surf} = N_{atm} \times E_k$. This power density is of the order of some W/cm$^2$ in the case of SCBD thus not producing an increase of the substrate temperature above RT. This should be compared with the power density for ion beams which is typically of the order of tenths of W/cm$^2$ [10, 17].

As demonstrated in a previous work on PMMA, Pd clusters penetrate inside the polymer matrix forming a polymer-metal nanocomposite [13]. The implanted clusters form a 3D network with a density increasing with the exposure time and eventually producing an electrically conducting region starting from several tens of nanometers beneath the polymer surface and emerging above the surface as a nanostructured metallic film (figure 1b).

The substrate to be implanted is mounted on a sample holder provided with a quartz microbalance in order to continuously monitor the cluster deposition rate [13]. Since the supersonic beam covers a circular spot with a diameter of 4 cm, in order to implant the surface of a 4-inch wafer we rastered the cluster beam by moving the substrate with respect to the beam axis along parallel vertical lines (figure 1c) in steps of 5mm. Partial overlapping of the stripes helps achieving a homogeneous thickness over large areas (see below), the rastering pattern is repeated until the desired metal equivalent thickness has been deposited (figure 1c).

## 3. Micropattern fabrication and characterization

*3.1 Fabrication procedure*

Before producing the final MEA devices, a microfabrication approach, sketched in figure 2, was planned in order to develop preliminary test structures (a set of microresistors), for evaluating the performances of the main technological steps and the morphological and electrical properties of the nanocomposite micropatterns. The process outline starts from (figure 2a) a blank 4" mono crystalline p-type (100) silicon wafer (thickness 500µm) where a layer of silicon oxide (300 nm) was grown onto the silicon substrate as a first passivation layer (figure 2b). Then a thick multilayer made by 1.2 µm of sacrificial positive photo resist (Fujifilm HiPr 6500 series), 100 µm Sylgard 184 (PDMS) and 140µm SU-8 3050 was spun and polymerized on the Silicon Oxide (figure 2c). In order to promote the lithographic definition of the metal layer, a further layer of MAN 1420 (Micro resist technology GmbH) negative photoresist was spun, exposed and developed to fabricate a lift-off mask (figure 2d). As following step, a layer of nanostructured palladium (thickness 10-25 nm) was deposited by SCBD (figure 2e). The Pd structure was then defined using a lift-off procedure (figure 2f). Final results are reported in Figs. 3a, 3b and 3c, showing patterns with minimum lateral dimension of 20 µm for the Pd layer. Another set of microresistors was produced following the previous method until step "*e*": in this case, a further layer of gold was subsequently deposited by thermal evaporation (figure 2e'). Finally, the Pd/Au structure was defined using the same lift-off procedure (figure 2f').

*3.2 Electrical characterization*

We characterized the electrical conductivity of the Pd cluster-assembled microresistors described in the previous sections (produced following the sketch of figure 2 until step "*e*", without the evaporation of the further gold layer). Since the absolute resistances were in the order of tens of kilo-ohm, measurements were executed with a two-probe set up, neglecting contact resistances. Each resistor was tested in the range between 0 and 5 volts, increasing the voltage by steps of 50 mV. In figure 4 we show the I-V curve of a microresistor sample: as it can be clearly seen, the linear response of the device indicates an ohmic behaviour. We used *ex post* electrical characterization of these microresistors also as a tool to check the reproducibility of the cluster-assembled Pd film thickness and the effect of supersonic cluster beam rastering. In particular, we produced complete arrays of microresistors on the same wafer (figure 1c), and then measured their sheet resistances. Performing statistical calculation, we found sheet resistances to have a standard deviation of 3.3% across the whole wafer surface. These measurements show the good reproducibility of the cluster-assembled Pd film thickness and the reliability of the technique previously described.

When microdevices with higher electrical conductance are needed it is possible to use a two-step process in order to improve the electrical performances: once the nanocomposite film is synthesized by SCBD technique, a further conductive layer of gold can be deposited using standard metallization techniques such as electron beam or thermal evaporation (figure 2e'). In this case, the Pd nanocomposite film acts as an adhesion layer for another, more conductive, film. The main advantage of this solution is a better adhesion between the metal layer and the polymer substrate, guaranteeing a higher stress resistance compared to the one obtained using standard thermal evaporation, as we will show in the following section.

*3.3 Adhesion*

To test the adhesion of the cluster-assembled metallic layer and to demonstrate its use as an effective adhesion layer for further metallization, we carried out a Scotch® tape test on microresistors produced by Pd nanoparticle implantation and subsequent deposition of a gold overlayer with different thicknesses as described in section 3.1. We fabricated two groups of 4 gold microresistors depositing a gold film by thermal evaporation with a thickness of 200 nm and 400 nm respectively (figure 5a); two resistors of each group had a nanocomposite Pd adhesion layer obtained by SCBD prior to gold evaporation.

Before performing the Scotch® tape test, the microresistors were sonicated in an acetone bath for ten minutes to remove the resist mask. Figure 5b and 5c show respectively the effect on the as-deposited microresistors of acetone sonication and of the Scotch® tape test: one can easily see how the gold microresistors without the nanocomposite adhesion layer are already significantly damaged by the sonication process while the gold microresistors with the ns-Pd/SU-8 adhesion layer demonstrate a remarkably good adhesion under the Scotch® tape test.

**4. Micro electrode array fabrication**

After the development and characterization of the test microresistors, a second microfabrication process was implemented with additional technological steps in order to develop the final device prototypes. In this aim, a new design specifically devoted to microelectrodes arrays (MEAs) for in-vitro growth and electrical stimulation of neural cells was developed. The additional steps allow the passivation of interconnection metals and the release of the polymeric-based thin substrate. For implementing this task a thin layer of SU8 3010 (Microchem) negative photo resist (6-10µm) was deposited and patterned to guarantee an adequate passivation for the interconnection wires of the microelectrodes (figure 2g and 2g'). The sacrificial positive resist layer was then removed by acetone (figure 2h and 2h'). Figure 6a shows the 4" wafer plan with the MEA (4 cm x 4 cm) final layout (in the center of the wafer, see figure 6b). Figure 6c shows a detail of a single passivated electrode having an exposed pad at its end. The mask was designed to get various test structures onto the full wafer surface, beside the central MEA. The photograph in figure 7 illustrates the obtained flexible SU-8 based device.

**5. Biocompatibility and cell culture test**

The biocompatibility test of the MEA and, in general, of the Pd/SU-8 nanocomposite was performed using PC-12 cell line (Rat Pheochromocytoma, ATCC). Cells were cultured in F12K medium culture (containing 2.5% fetal bovine serum; 15% horse serum 100 U/ml penicillin and 100 µg/ml streptomycin; Sigma Aldrich) according to standard procedures. Cells were passaged every 3-5 days and used within 35 passages. Cells were detached and disaggregated in solution with a micropipette and plated on MEA ($2 \cdot 10^6$ cells/MEA). After 4 hours, cells were washed three times with PBS to remove unattached cells and were incubated for 24-48-72 hours for further analysis by adding fresh culture media. Cell adhesions were monitored and images were taken with a CCD camera mounted on a Zeiss Axio Imager.A1m microscope by using an 10x objective lens.

The biocompatibility of SU-8 has been reported [18,19], however *in vitro* tests on neural cells are still scarce and not systematic [20]. We have performed a comparative *in vitro* characterization using substrates where different regions have been modified by Pd cluster implantation and subsequent gold films deposition. In figure 8a we show cells grown on 3 adjacent substrates: bare SU-8, Pd/SU-8 nanocomposite and gold-plated Pd/SU-8 nanocomposite. One can see how cells prefer to adhere on the Pd/SU-8 nanocomposite (both simple and gold-plated) than to the pure SU-8 substrate. Moreover the picture points out a higher cell density on the half plate coated with gold. We found that the cell density on the gold-plated Pd/SU-8 nanocomposite is 1.8 times higher than the density on the simple Pd/SU-8 nanocomposite. This observation could be due either to an increased cell adhesion on the gold-plated nanocomposite, or to an increased proliferation of cells, always on the gold-plated area.

The preference for ns-Pd/SU-8 regions (both gold-coated and not) is also made apparent by cultivating PC12 on MEA: Figs 8b and 8c show the preferential growth of cells on the electrode regions, whereas the bare SU-8 substrates does not support cell growth and adhesion. We point out that no any kind of surface functionalization has been performed, neither on the bare SU-8 nor on the metallized surfaces. This is particularly interesting for the use of MEAs in *in vitro* experiments due to the spontaneous adhesion of cells only on the electrodes. Moreover, since no functionalization process is involved, the adhesion properties are not prone to possible deteriorations of the functional molecule.

The observed increase in biocompatibility of ns-Pd/SU-8 surfaces may be attributed to both a variation in the surface roughness due to cluster implantation and to an increased stiffness of the metallized region with respect to bare SU-8. Indeed, on the one hand surface morphologies on the nanometric scale strongly influence protein adsorption on surface and cell adhesion [21], while on the other hand neuronal cells (as many other cell types) are sensitive to the mechanical properties of their extracellular matrix (ECM), such as the substrate stiffness [22].

## 6. Conclusions

Supersonic cluster beam deposition on SU-8 has been used to fabricate microresistors and microelectrode arrays. We produced conductive Pd/SU-8 nanocomposite with good and reproducible electrical conduction properties which can also be used as adhesion layers for metallic films deposited with traditional deposition techniques. Unlike in ion implantation processes, in SCBD metallic clusters are formed before implantation into the polymeric matrix.

We showed that SCBD is compatible with standard lift-off technology on 4'' wafers, and we developed a complete process for producing polymer based MEAs which are self standing, transparent, flexible, and biocompatible. The nanocomposite showed an enhanced biocompatibility for PC12 cell growth and adhesion, resulting in a spontaneous confinement of cells on the nanocomposite areas.

**Acknowledgements**

This work has been partially supported by Fondazione CARIPLO under the project ''Sviluppo di sistemi di cultura cellulare su materiali biocompatibili nanostrutturati per lo studio di patologie a scopo eziologico e terapeutico''.

Flexible and biocompatible microelectrode arrays fabricated by SCBD on SU-8

Flexible and biocompatible microelectrode arrays fabricated by SCBD on SU-8

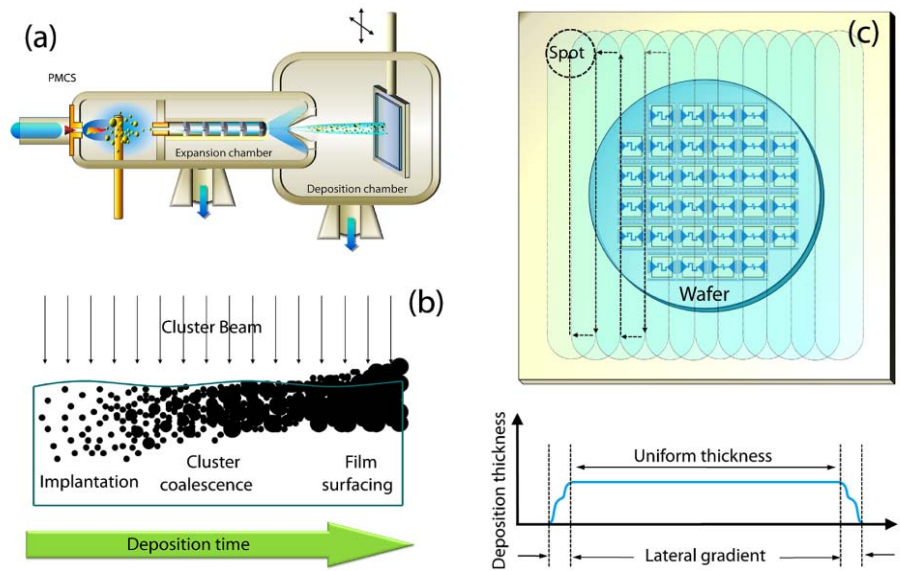

**Figure 1**: (a) Schematic view of the SCBD apparatus (not to scale). (b) Schematic representation of the growth of metal/polymer nanocomposite resulting from individual cluster implantation, cluster coalescence and cluster percolation to form a continuous film. (c) Rastering process during cluster beam deposition. A wafer with microresistors is represented. The rastering process produces vertical metallized stripes – partially overlapped with one another – yielding a central area with a homogeneous metal thickness, whereas the periphery is characterized by thickness gradients.

Flexible and biocompatible microelectrode arrays fabricated by SCBD on SU-8

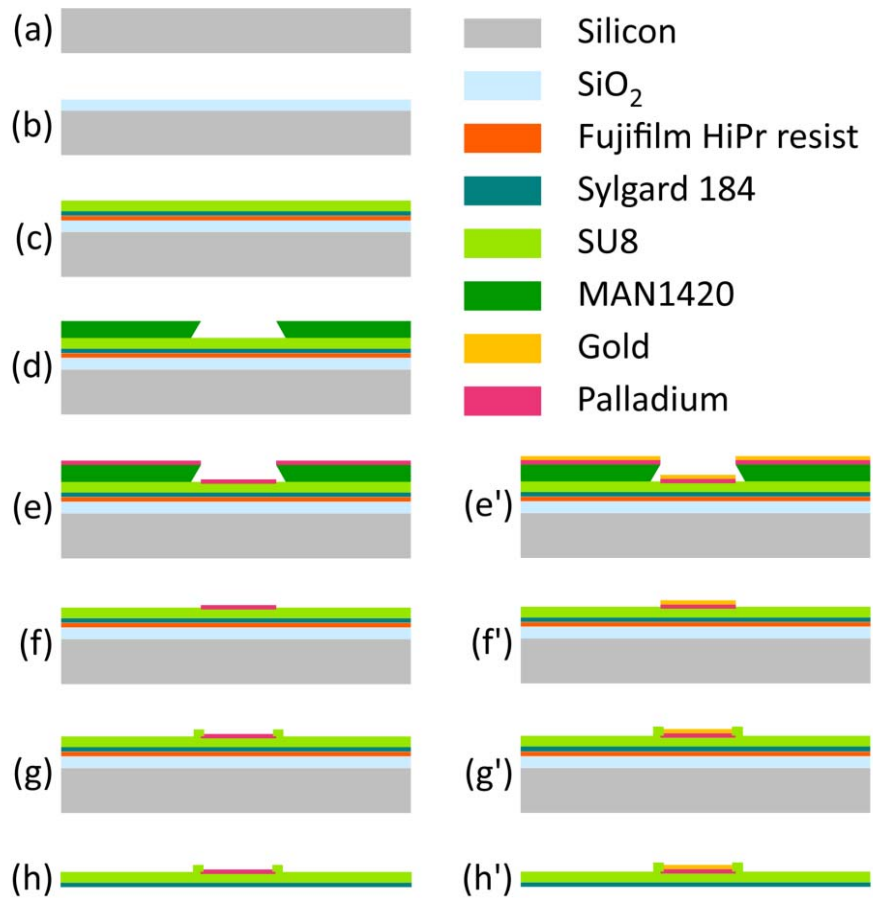

**Figure 2**: Microresistors deposition process: a Si wafer (a) with a 300nm thick silicone oxide layer (b) is used as substrate to spin coat sequentially three polymeric layers: Fujifilm HiPr (6500 series), Sylgard 184 and SU-8 (c). Then a resist mask (d) for microresistors definition is obtained by photolithography. A layer of nanostructured palladium is deposited by SCBD, leading to the formation of the Pd/SU-8 nanocomposite (e). An additional layer of gold can subsequently be deposited by electron beam or thermal evaporation (e'). The photresist mask is removed with a lift-off step in order to define the Au/ns-Pd structure (f and f'). Further steps can include the passivation of interconnection metals (g and g') and the release of the polymeric-based thin substrate (h and h').

Flexible and biocompatible microelectrode arrays fabricated by SCBD on SU-8

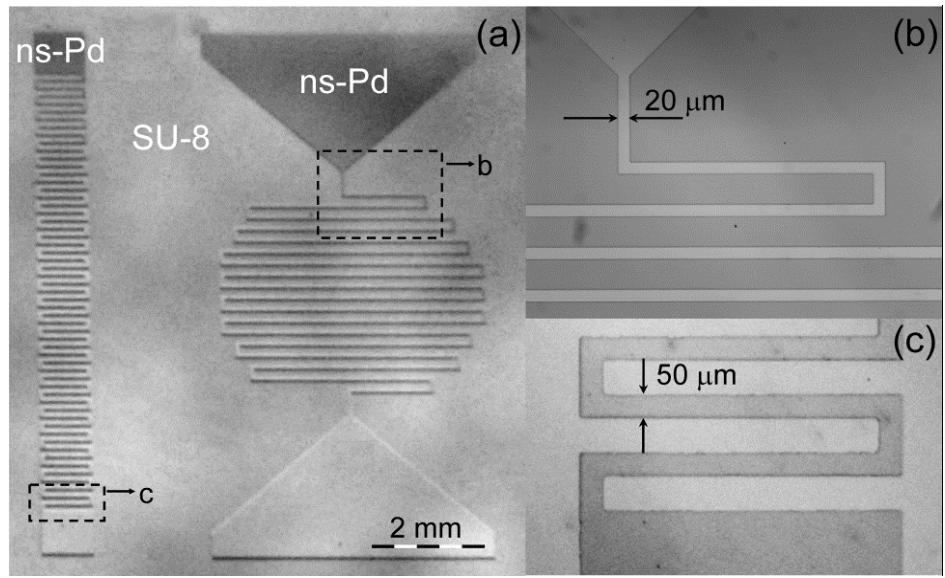

**Figure 3**: (a) Optical image of microresistor fabricated by metallization with Pd nanoparticles. Details of the microresistor are reported in (b) and (c). Patterns with feature sizes down to 20 μm can be easily achieved.

Flexible and biocompatible microelectrode arrays fabricated by SCBD on SU-8

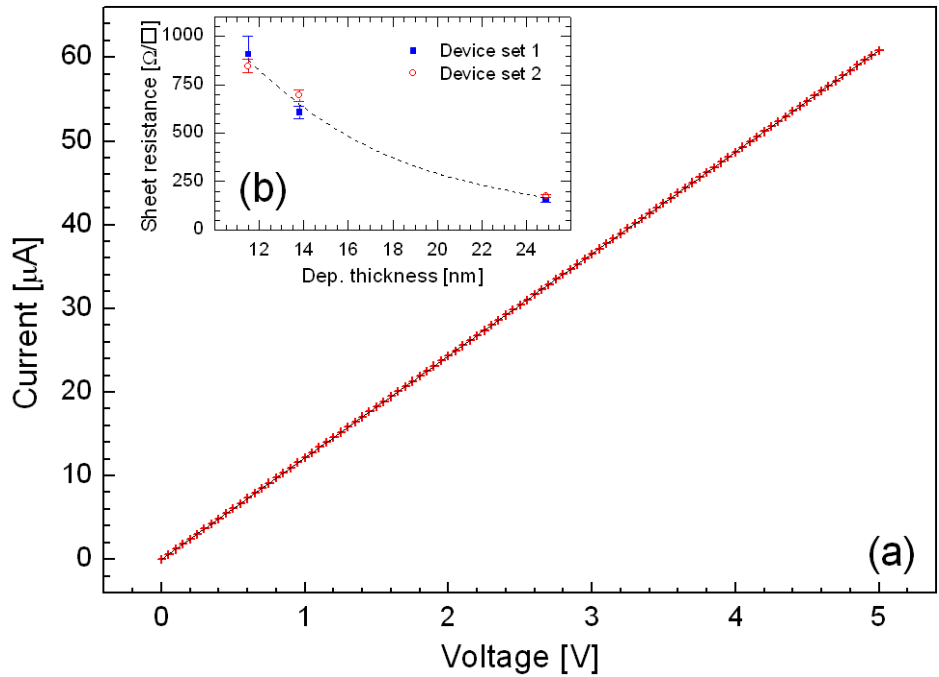

**Figure 4**: (a) Experimental I-V curve from the two-probes measurement of a Pd cluster-assembled microresistor. The linear fit is indicated as a black line. In the inset (b) the sheet resistance values achieved on three set of microresistors with different deposition thickness of Pd are shown.

Flexible and biocompatible microelectrode arrays fabricated by SCBD on SU-8

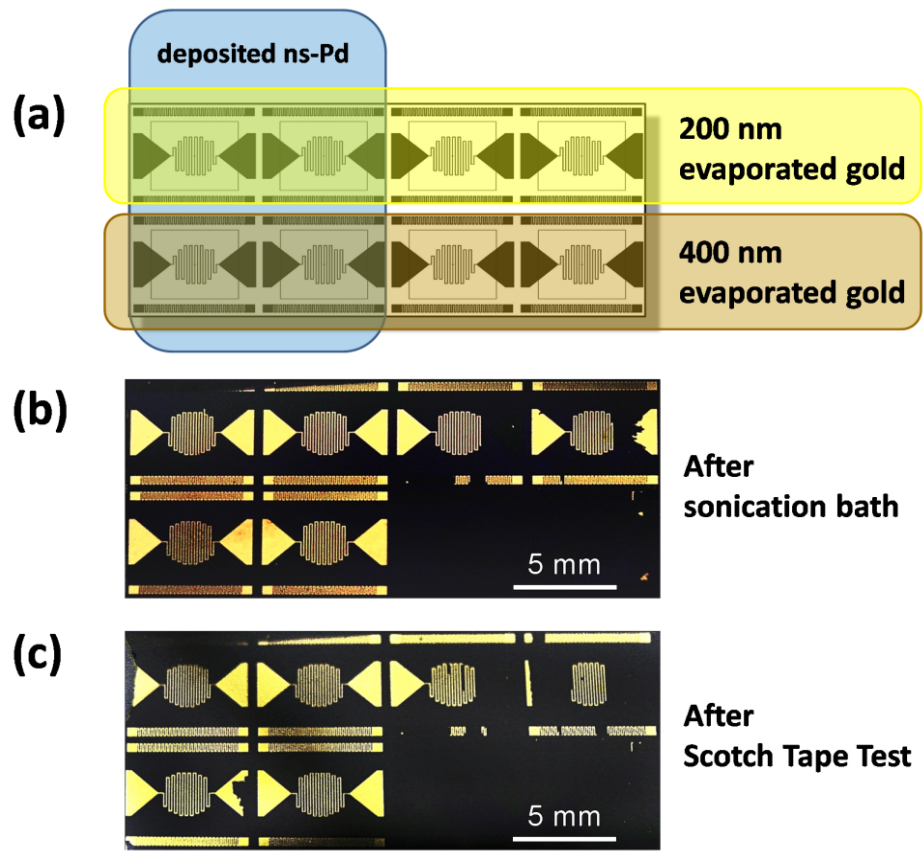

**Figure 5**: (a) Au microresistors preparation: the microresistors are made as follows: **top-left**: 200 nm gold film evaporated onto a Pd/SU-8 nanocomposite; **top-right**: 200 nm gold film evaporated onto simple SU-8; **bottom-left**: 400 nm gold film evaporated onto a Pd/SU-8 nanocomposite; **bottom-right**: 400 nm gold film evaporated onto simple SU-8 (the resistor is completely removed by the ultrasonic bath in acetone).(b) A sample of Au micro resistors prepared as described in (a) just after the acetone sonication bath. (c) The same micro resistors after a Scotch® tape test.

Flexible and biocompatible microelectrode arrays fabricated by SCBD on SU-8

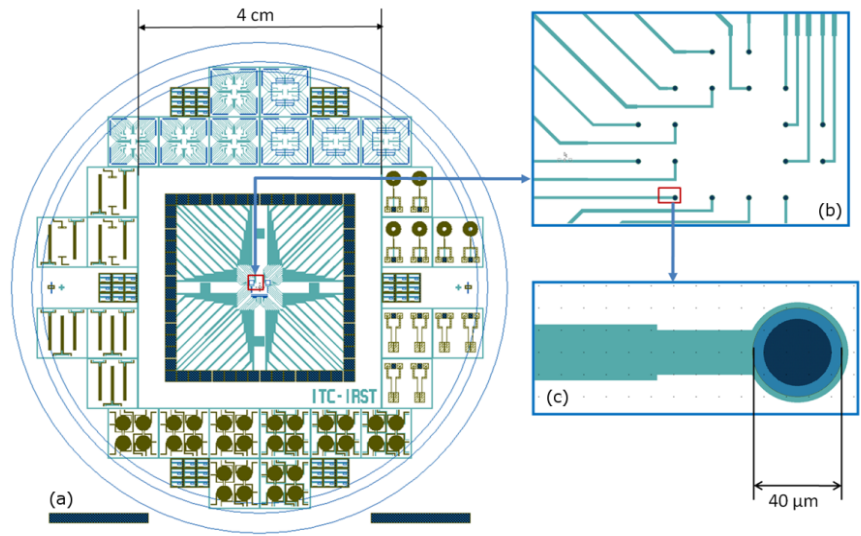

Figure 6: (a) The layout used for the production of the 4X4 cm$^2$ MEA on a 4" wafer. (b) The central part of the MEA, with micrometrical features. (c) The detail of a single electrode, showing the exposed pad after passivation.

Flexible and biocompatible microelectrode arrays fabricated by SCBD on SU-8

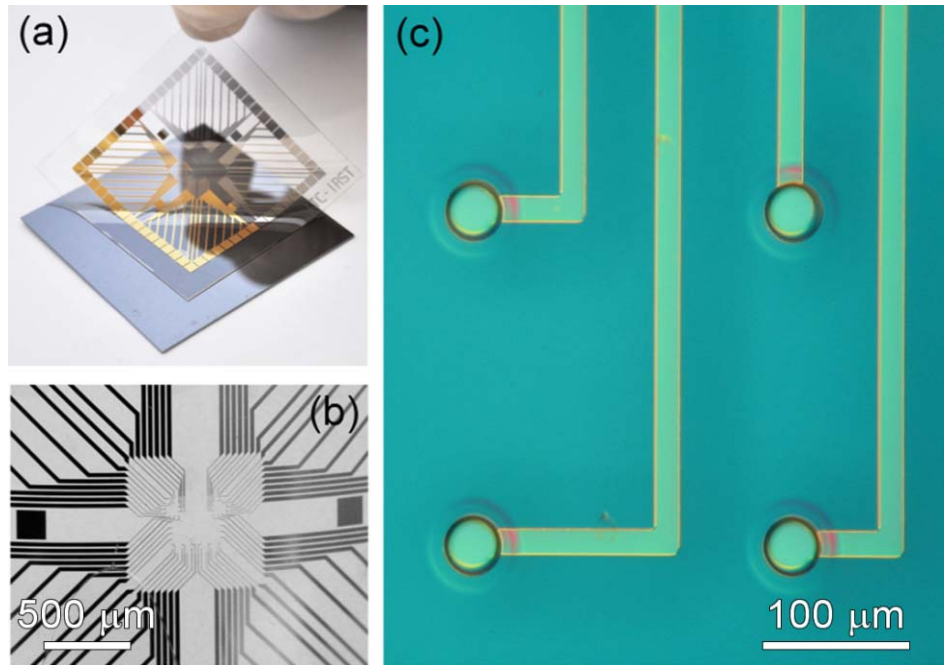

**Figure 7**: (a) Self standing flexible SU-8 based MEA detached from the Si substrate upon removal of the sacrificial layer (see figure 2h). On the right half of the device the microelectrodes are made by Pd/SU-8 nanocomposite (see figure 2e), while on the left half a further gold layer was evaporated on the Pd/SU-8 nanocomposite (see figure 2e'). (b) and (c) show higher magnification of MEA details.

Flexible and biocompatible microelectrode arrays fabricated by SCBD on SU-8

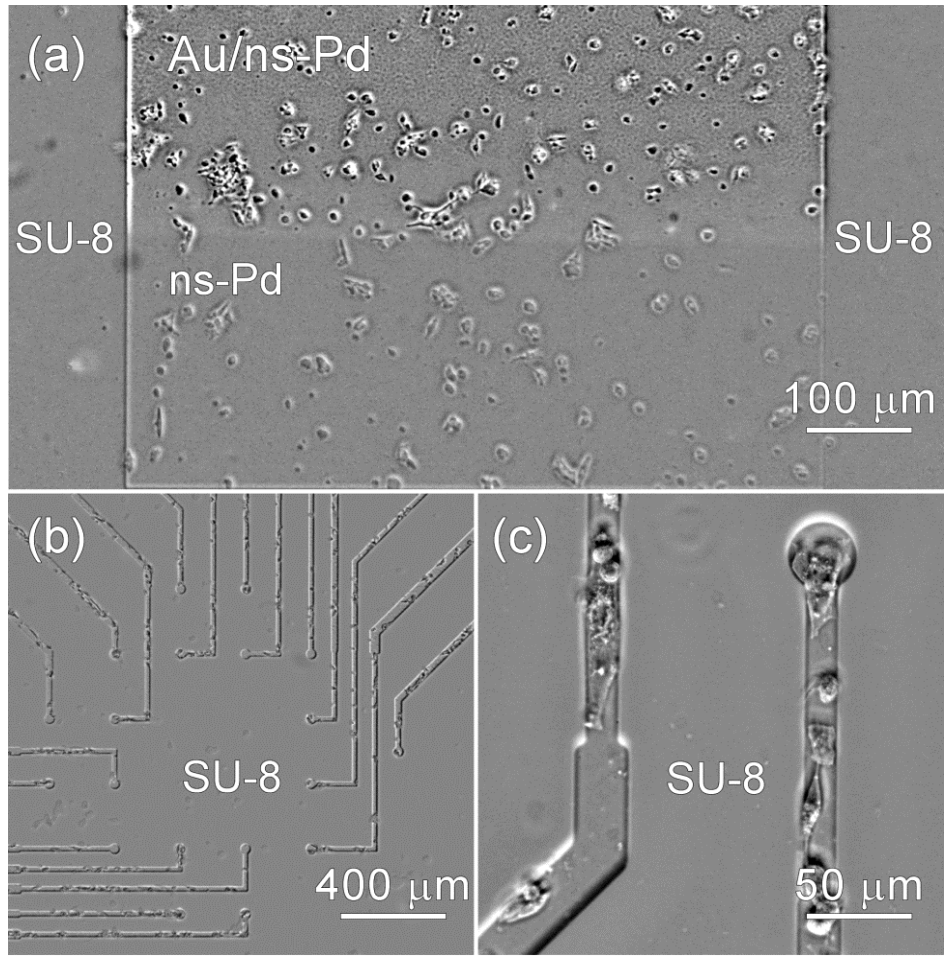

**Figure 8**: PC-12 cells on different parts of the MEA. (a) A micrograph of a metallized plate surrounded by unimplanted SU-8. The lower part of the plate is made of Pd/SU-8 nanocomposite (ns-Pd), while the upper part was made by evaporation of a gold film over the Pd/SU-8 nanocomposite (Au/ns-Pd). The cell density on gold-coated Pd/SU-8 nanocomposite is roughly 1.8 times higher than the cell density on the non-coated half. (b) and (c) details of different parts of the electrodes of the MEA device showing selective cells adhesion on the metallized areas.